\documentclass[a4paper,fleqn,usenatbib]{mnras}

\usepackage{newtxtext,newtxmath}

\usepackage[T1]{fontenc}
\usepackage{ae,aecompl}

\usepackage{graphicx}	
\usepackage{amsmath}	
\usepackage{amssymb}	
\usepackage{bm}

\title[Rotation and magnetism in intermediate mass stars]{Rotation and magnetism in intermediate mass stars}

\author[L.\,G.\,Quentin and C.\,A.\,Tout]{
L\'eo G. Quentin,$^{1}$\thanks{E-mail: leo.quentin@ens.fr}
Christopher A. Tout$^{2}$
\\
$^{1}$D\'epartement de physique, \'Ecole normale sup\'erieure,
45 rue d'Ulm, 75005 Paris, France\\
$^{2}$Institute of Astronomy, University of Cambridge, The Observatories, Madingley Road, Cambridge CB3 0HA\\
}

\date{Accepted XXX. Received YYY; in original form ZZZ}

\pubyear{2017}

\begin{document}
\label{firstpage}
\pagerange{\pageref{firstpage}--\pageref{lastpage}}
\maketitle

\begin{abstract}
Rotation and magnetism are increasingly recognized as important
phenomena in stellar evolution.  Surface magnetic fields from a few to
$20{,}000\,$G have been observed and models have suggested that
magnetohydrodynamic transport of angular momentum and chemical
composition could explain the peculiar composition of some stars.
Stellar remnants such as white dwarfs have been observed with fields
from a few to more than $10^{9}\,$G.  We investigate the origin of and
the evolution, on thermal and nuclear rather than dynamical
time-scales, of an averaged large-scale magnetic field throughout a
star's life and its coupling to stellar rotation.  Large-scale
magnetic fields sustained until late stages of stellar evolution with
conservation of magnetic flux could explain the very high fields
observed in white dwarfs.  We include these effects in the Cambridge
stellar evolution code using three time-dependant advection-diffusion
equations coupled to the structural and composition equations of stars
to model the evolution of angular momentum and the two components of
the magnetic field.  We present the evolution in various cases for a
$3\rm\,M_{\odot}$ star from the beginning to the late stages of its
life.  Our particular model assumes that turbulent motions, including
convection, favour small-scale field at the expense of large-scale
field.  As a result the large-scale field concentrates in radiative
zones of the star and so is exchanged between the core and the
envelope of the star as it evolves.  The field is sustained until the
end of the asymptotic giant branch, when it concentrates in the
degenerate core.
\end{abstract}

\begin{keywords}
  stars: evolution, stars: general, stars: magnetic fields, stars: rotation, (stars:) white dwarfs
\end{keywords}



\section{Introduction}

While rotation and magnetism have often been considered as minor
effects in stellar evolution, observations have shown that
these phenomena must be taken into account when describing a star's
life.  Let us first briefly present the main characteristics of these
effects in stars and how magnetic fields in white dwarfs could be
linked to the evolution of magnetic fields through stellar evolution
processes.

Rotation arises from conservation of angular momentum.  Stars form when
giant clouds of gas undergo gravitational collapse.  Because such
clouds have large-scale turbulence, different fluid particles have
different velocities, resulting in a non-zero total angular
momentum.  Therefore, as a cloud collapses and contracts, its material
sees its rotation rate increase greatly.  When stars eventually form, it
is highly likely that they spin around a particular axis.  Because of
the random nature of turbulence, stars can be found with various
angular velocities.  So, owing to the centrifugal force, rotating
stars see their hydrostatic balance modified and their structure
changed.  This is very similar to what can be observed on Earth, where
the equatorial radius is approximately $21\,\rm km$ larger than the
polar radius.  Alongside these structural changes, other effects such
as rotationally-driven turbulent mixing or meridional circulation,
that is to say from one pole to the other, can also deeply affect the
star's chemical evolution.  Rotation in stars is very different from
Earth's rotation because stars do not necessarily rotate as solid
bodies such that material in a star can have different rotation
rates at different latitudes and depths, leading to shears.  These
shears are of particular importance in the development of the
instabilities that drive additional mixing.

Stellar magnetism is also a phenomenon that can be observed within our
close neighbourhood.  The Sun's surface field has a strength of, on
average, $1\,$G.  Observations of other stars have shown that the
intensity of stellar magnetic fields covers many decades.  Some, of
type Ap and~Bp, have surface fields ranging from $100$ to
$20{,}000\,$G and show chemically peculiar surface compositions
compared to normal A~type stars.  It has therefore been suggested that
the action of magnetic fields has an influence on the composition of
these stars.  Some elements that are usually present only in the inner
parts of stars are radiatively levitated.  In short, the strong
surface magnetic field stabilizes against turbulence, so that
levitation concentrates the heavy elements at the photosphere.  Very
high magnetic fields are also common features in compact objects.
White dwarfs are found with fields ranging between zero to~$10^{9}\,$G
and some magnetic neutron stars can have field strengths of
order~$10^{15}\,$G.  So magnetic effects are present and can be
important throughout the life of most stars.

Magnetic fields in stars show very complex behaviour, with structure on small and
large scales.  The origin of the large-scale fields is
still debated.  For stars on their main sequence, the phase of their
evolution where hydrogen is fused in the core, two possibilities are
often considered.  They could either be fossil fields, that can be dated
back to the star's formation, or fields generated by a dynamo effect, as is
thought to be the case in the Sun.  In the first case, weak fields
within a protostellar gas cloud can become enhanced through flux
conservation as the material collapses, giving birth to more intense
fields.  In the second case, energy is transferred between the kinetic
motion of the material in the star to the magnetic field.  Large-scale
magnetic fields in stars can be unstable, so dynamo effects are highly
interesting because they can regenerate these fields.

Hypotheses for the origin of magnetic fields in white dwarfs are the
same as those for main-sequence stars.  Therefore we wish to know if
large-scale fields can be sustained throughout the life of a star,
whether they are fossil fields or dynamo generated.  White dwarfs are
stellar remnants of intermediate- and low-mass stars.  At the end of
their lives, these stars have completely exhausted the fuel they can
burn in their core, which collapses until an equilibrium between
gravity and electron degeneracy pressure is reached.  White dwarfs emerge
as the outer layers of stars are shed by various processes,
leaving only the degenerate core.  Very high magnetic fields, more than
$10^{7}\,$G, have been observed in white dwarfs, so one of our
motivations is to see if we can obtain such high fields through the
evolution of intermediate-mass stars.

We begin by presenting main features of the Cambridge stellar
evolution code (STARS) that we use for numerical simulation of stellar
evolution, including the model used to describe how both rotation and
magnetism have been incorporated in the code in its new version
Rotating Stellar Evolution (RoSE) and how exactly we have modified it.
We then use the code to derive some results concerning the evolution
of the magnetic field in a $3\,\rm M_{\odot}$ star for our particular
magnetohydrodynamic model.  Finally, we briefly present how the evolution of such
fields can hint at the origin of the very high magnetic field observed
in some white dwarfs.

Most of the theoretical framework in the next sections is based on the
dissertation of \cite{PotterThesis} and follows the work of
\citet*{PotterToutEldridge2012a}, \citet*{PotterToutBrott2012b},
\citet*{PotterChitreTout2012c} and the book \textit{physics, formation
  and evolution of rotating stars}, by \cite{Maeder2009}.  The first
version of RoSE was written during the PhD of A. Potter in 2012 to
model rotation and magnetic field evolution in massive stars.  Since
then, we have modified it to be more stable, especially for low-mass
stars and during the late phases of evolution.

The code is based on the Cambridge stellar evolution code, which we
refer to as STARS.  This code was originally developed by
\cite{Eggleton1971} and has been revised and improved during the past
40\,yr \citep[see in particular][]{Pols1995,StancliffeEldridge2009}.
It is written in FORTRAN~77 and is approximately 7000~lines long.
Details of the latest stellar physics are given by
\cite{StancliffeEldridge2009}.  Succinctly, the code computes the
stellar structure and chemical evolution simultaneously with a
Newton--Raphson algorithm.

The structure and composition of the modelled star are calculated on a
one-dimensional sequence of mesh points.  The mesh is non-Lagrangian
and the points are distributed uniformly in a spacing~$Q$, a
rather complicated function of pressure, temperature, mass, radius and
density of the star.  This arranges the mesh points so that higher
resolution can be obtained where it is needed, such as near the
burning shells in giants.  This mesh-spacing function can quite easily
be modified when new features are added.

The RoSE code includes additional equations for the evolution in one
dimension of the effective radius $r$ within the star, angular
velocity $\Omega(r)$ and two components of the large-scale magnetic
field $B_{\rm p}(r)$, representing an averaged poloidal filed, and
included by a magnetic potential $A(r)$, and $B_\phi$, representing an
averaged toroidal field.  Because the STARS code is one-dimensional we
cannot model angular variations of either $\Omega$ or $\pmb{B}$.  This
is certainly a significant approximation but one that remains
necessary until full stellar evolution in two or three dimensions is
possible.  We also want to follow evolution over the nuclear lifetime
of the star and so cannot model short term variations in either
$\Omega$ or $\pmb{B}$.  This excludes any form of dynamo cycle from
our models.  We therefore require that $\Omega$, $A$ and $B_\phi$
remain positive throughout the evolution and so represent an average
magnitude over any small-scale, short-term evolution.


\section{Rotating stars}

We know that, owing to its rotation, the Earth is not a sphere but is
slightly distorted.  Its equatorial radius is about $21\,$km longer
than its polar radius.  Relatively, some stars rotate much faster.  In
the most extreme cases the equatorial radius of a star can be up to
about 1.5~times the polar radius. This deformation is caused by the
centrifugal force, that also affects the internal hydrostatic balance
of the star.  Furthermore, helioseismology has shown that the sun does
not rotate as a solid body but undergoes differential
rotation.  Therefore, our model assumes that the angular velocity
within the interior of a star can vary.  Both of these properties of
stellar rotation induce internal circulation motions and instabilities
that transport chemical elements and angular momentum.  In this
section, we describe how we modify the previous model to include the
effects of rotation.

\subsection{Structure equations}

An important hypothesis on which we rely to derive the structure and
evolution equation comes from \citet{Zahn1992}.  He first assumes that
stars undergo differential rotation.  Because of this, he argues that the
shear-driven horizontal turbulence is much stronger than the
vertical turbulence.  Therefore variations in physical properties are
much smaller horizontally than vertically and the angular velocity is
assumed to be constant on isobaric shells.  This approximation is
referred to as the \textit{shellular hypothesis}.  It also applies to
the state variables and composition variables.  This
approximation is necessary to us because it reduces a
two-dimensional problem to one dimension, allowing us to
easily adapt the STARS code.  Let us now describe the modified
equations that we use.

We define $S_{\!p}$ to be a surface of constant pressure $P$, the volume
contained within $S_{\!p}$ to be $V_{\!p}$, $r_{\!p}$ to be the radius of a
sphere with volume $V_{\!p}$ and $m_{\!p}$ the mass enclosed within
$S_{\!p}$.  With these variables the mass conservation equation keeps
its standard form
\begin{equation}
\frac{\text{d} m_{\!p}}{\text{d}r_{\!p}}=4\pi r^{2}_{\!P}\rho,
\end{equation}
where $\rho$ is the density on the isobaric surface.

In a rotating star the local gravity vector is
\begin{equation}
\bm{g}_{\text{eff}}=\left ( - \frac{Gm_{\!p}}{r^{2}}+\Omega^{2} r \sin^{2} \theta \right ) \bm{e}_{r} + \left ( \Omega^{2} r \sin \theta \cos \theta \right ) \bm{e}_{\theta},
\end{equation}
where $\Omega$ is the local angular velocity, $G$ is the gravitational
constant, $\theta$ is the angle from the polar direction and
$\bm{e}_{r}$ and $\bm{e}_{\theta}$ are the radial and spherical polar
unit vectors.  We have neglected the effect of the aspheric mass
distribution on gravity because stars are generally centrally
condensed and the approximation that gravity, owing to mass, acts
radially is good.  To obtain one-dimensional
equations, we define the average of any quantity $q$ over $S_{\!p}$ as
\begin{equation}
< q > = \frac{1}{S_{\!p}} \int_{S_{\!p}} q \text{d} \sigma,
\end{equation}
where $\text{d}\sigma$ is a surface element of $S_{\!p}$.  With this
notation, and according to the derivation of \cite{Maeder2009}, the
equation of hydrostatic equilibrium becomes
\begin{equation}
\frac{\text{d} P}{\text{d} m_{\!p}}= - \frac{G m_{\!p}}{4 \pi r^{4}_{\!p} } f_{\!p},
\end{equation}
where $P$ is the pressure and 
\begin{equation}
f_{\!p}=\frac{4\pi r^{4}_{\!p}}{G m_{\!p} S_{\!p}} < g_{\text{eff}}^{-1} >^{-1}.
\end{equation}
That is the standard hydrostatic equation is adjusted by a factor $f_{\!p}$
which tends to unity for non-rotating stars.

In a similar way the equation for radiative equilibrium is modified to 
\begin{equation}
\frac{\text{d} \log T}{\text{d} \log P} =\frac{3 \kappa P L_{\!p}}{16 \pi a c G m_{\!p} T^{4}} \frac{f_{T}}{f_{\!p}},
\end{equation}
where $T$ is the temperature, $\kappa$ is the opacity, $a$ is the
radiation constant, $c$ is the speed of light, $L_{\!p}$ is the total
energy flux through $S_{\!p}$ and
\begin{equation}
f_{T}= \left ( \frac{ 4\pi r_{\!p}^{2}}{S_{\!p}}  \right ) \left ( < g_{\text{eff}} > < g_{\text{eff}}^{-1} > \right )^{-1},
\end{equation}
again the same as in the non-rotating case except for the
multiplication by $f_{T}/f_{\!p}$, which also tends to unity as the
rotation vanishes.

In a non-rotating star in hydrostatic equilibrium, the
transport of angular momentum and of chemical elements would be purely
diffusive.  However, according to von Zeipel's theorem
\citeyearpar{vonzeipel1924}, the thermal flux $F$ through a point in a
rotating star behaves as $F \propto g_{\text{eff}}(\theta)$.  Because
$g_{\text{eff}}$ depends strongly on co-latitude $\theta$ there is a
thermal imbalance that drives a meridional circulation.  We introduce the
meridional circulation of \cite{MaederZahn1998}
\begin{equation}
\label{eq:meridionalvelocity}
\bm{U}=U(r) P_{2}(\cos \theta) \bm{e}_{r}+V(r) \frac{\text{d} P_{2}( \cos \theta)}{\text{d} \theta} \textbf{e}_{\theta},
\end{equation}
where $U$ and $V$ are linked by the continuity equation so that 
\begin{equation}
V=\frac{1}{6 \rho r} \frac{\text{d}}{\text{d} r} ( \rho r^{2} U )
\end{equation}
and $P_{2}$ is the second Legendre polynomial
$P_{2}(x)=\frac{1}{2}(3x^{2}-1)$.  An approximate expression for $U$
is
\begin{equation}
U=C_{0} \frac{L}{m_{\text{eff}} g_{\text{eff}}}\frac{P}{C_{\!p}\rho T} \frac{1}{\nabla_{\text{ad}}-\nabla_{r} +\nabla_{\mu}} \left (1 - \frac{\epsilon}{\epsilon_{m}} - \frac{\Omega^{2}}{2 \pi G \rho} \right ) \left ( \frac{ 4 \pi^{2} r^{3}}{3 G m } \right ),
\end{equation}
where $L$ is the stellar luminosity, $C_{0}$ is a constant used for
calibration, $m_{\text{eff}}= m \left( 1- \frac{\Omega^{2}}{2\pi G
  \rho }\right)$, $\epsilon=E_{\text{nuc}}+E_{\text{grav}}$ is the
total local energy emission, $\epsilon_{m}=L/m$, $C_{\!p}$ is the
specific heat capacity at constant pressure, $\nabla_{\text{ad}}$ and
$\nabla_{\text{r}}$ are the adiabatic and radiative temperature
gradients and $\nabla_{\mu}=\frac{\text{d} \log \mu}{\text{d} \log P}$
is the mean molecular weight gradient.

We expect enhanced mass loss owing to
rotation.  \citet{FriendAbbott1986} suggested that near-critical
rotation drives mass loss to remove angular momentum and prevents the
surface from reaching critical velocity.  For a star of mass $M$ and
radius $R$ we modify the not rotating mass-loss rate
$\dot{M}_{\Omega=0}$ of \cite{Reimers1075} according to
\citet{Langer1998},
\begin{equation}
\label{Equ:MassLoss}
\dot{M}=\dot{M}_{\Omega=0} \left ( \frac{1}{1- \frac{\Omega}{\Omega_{\text{crit}}} } \right )^{\xi},
\end{equation}
where $\Omega_{\text{crit}}=\sqrt{\frac{2GM}{3R^{3}}}$ and $\xi=0.45$.  
In subsequent sections, we drop the suffix $p$ on $r_{\!p}$ and
$m_{\!p}$ for ease of reading.

\subsection{Angular momentum transport}

Shear creates stress, and therefore
dissipates energy, so a star tends towards its lowest energy state of solid body
rotation.  Because differential rotation is induced by, among other
factors, hydrostatic structural evolution, mass loss and meridional
circulation, we expect stars to be subject to a number of local
hydrodynamic instabilities.  The most important for us the
shear instability, which occurs when there is a velocity
difference between two layers of a fluid that have different
properties.  We assume that it dominates the vertical instabilities.
The instability drives turbulence that leads to diffusion of radial
and latitudinal variations in the angular velocity to bring the star
back to its lowest energy state.  This occurs with characteristic
diffusion coefficients $D_{\text{shear}}$, for the shear
instability acting vertically, and $D_{\text{h}}$ for horizontal
turbulence.  In the shellular hypothesis, we assume that the turbulent
mixing caused by these instabilities is much stronger horizontally
than vertically, so that $D_{\text{h}} \gg D_{\text{shear}}$.

Including all these processes and using Zahn's \citeyearpar{Zahn1992}
formulation, we obtain the diffusion--advection equation used to describe the
evolution of angular momentum
\begin{equation}
\label{eq:angularmomevolution}
\begin{aligned}
\frac{\partial (r^{2} \Omega ) }{\partial t} &= \frac{1}{5 \rho r^{2}} \frac{\partial (\rho r^{4} \Omega U)}{\partial r} + \frac{1}{\rho r^{2}} \frac{\partial}{\partial r} \left ( \rho D_{\text{shear}} r^{4} \frac{\partial \Omega}{\partial r} \right )\\
&+ \frac{1}{\rho r^{2}} \frac{\partial } { \partial r} \left ( \rho D_{\text{con}} r^{2+n} \frac{ \partial r^{2-n} \Omega}{\partial r} \right ),
\end{aligned}
\end{equation}
where $D_{\text{con}}$ is the convective diffusion coefficient and the
parameter $n$ is chosen to tend toward either solid body rotation or
uniform angular momentum in the convective zones when respectively
$n=2$ or $n=0$.  To obtain the boundary conditions for the angular
momentum, we integrate this equation over the star and get
\begin{equation}
\frac{\text{d} H_{\text{tot}}}{\text{d}t}=4 \pi r^{4} D_{\text{shear}} \rho \left. \frac{\partial \Omega}{\partial r} \right |^{R}_{0},
\end{equation} 
where $H_{\text{tot}}$ is the total angular momentum.  So, to ensure
angular momentum conservation, we require $\frac{\partial
  \Omega}{\partial r}=0$ at the surface and the centre. This condition
is modified in presence of magnetic braking.

The equations for the chemical evolution are 
\begin{equation}
\frac{\partial X_{i}}{\partial t}= \frac{1}{r^{2}} \frac{\partial}{\partial r} \left ( (D_{\text{shear}}+D_{\text{eff}}+D_{\Omega=0}+D_{\text{con}}) r^{2} \frac{\partial X_{i}}{\partial r} \right ),
\end{equation}
where $X_{i}$ is the mass fraction of element $i$.  We take
$D_{\text{con}}=(C_{\text{con}}(\nabla_{\text{r}}-\nabla_{\text{ad}})^{2}m^{2})/\tau_{\text{nuc}}$,
where $C_{\text{con}}$ is a large constant calculated from
mixing-length theory and $\tau_{\text{nuc}}$ is the nuclear time
scale.  The coefficient $D_{\text{eff}}$ describes the effective diffusion of chemical
elements owing to the interaction between horizontal diffusion and
meridional circulation.  Again according to \cite{Zahn1992}, we take
\begin{equation}
D_{\text{eff}}=\frac{\lvert r U\rvert^{2}}{30 D_{h}}.
\end{equation}
Finally, $D_{\Omega=0}$ is the diffusion coefficient for chemical
evolution without rotation.  It includes simple diffusion processes,
computed according to \cite{Paquette1986}, convective overshooting, as
included in the code by \citet*{Schroder1997} and semiconvection as
implemented by \cite{Eggleton1972}.  The coefficient $D_{\text{con}}$
is non-zero only in convective zones and $D_{\text{shear}}$ and
$D_{\text{eff}}$ are non-zero in radiative zones.  We use the
$D_{\text{shear}}$ of \cite{Talon1997},
\begin{equation}
D_{\text{shear}}=\frac{2 Ri_{\rm c} ( r \frac{ \text{d} \Omega}{\text{d} r} )^{2} }{N_{T}^{2}/(K+D_{\text{h}})+N_{\mu}^{2}/D_{\text{h}}},
\end{equation}
where $Ri_{\rm c}=(0.8836)^{2}/2$ is the critical Richardson number
\citep{Maeder2003}, $K$ is the radiative diffusivity and the
Brunt-V\"ais\"al\"a frequency has been split into
\begin{equation}
N_{T}^{2} = - \frac{g_{\text{eff}}}{H_{\!p}} \left ( \frac{\partial \ln \rho}{\partial \ln T} \right )_{P,\mu} ( \nabla_{\text{ad}}-\nabla),
\end{equation}
where $\nabla$ is the local temperature gradient, and
\begin{equation}
N_{\mu}^{2} = - \frac{g_{\text{eff}}}{H_{\!p}} \left ( \frac{\partial \ln \rho}{\partial \ln \mu} \right )_{P,T} \frac{\text{d} \ln \mu}{\text{d} \ln P},
\end{equation}
where $H_{\!p}$ is the pressure scale-height and $\mu$ is the chemical
potential.  Following \cite{Maeder2003}, we set the horizontal turbulence diffusion
coefficient to
\begin{equation}
D_{\text{h}}=0.134r(r\Omega V(2 V - \alpha U))^{1/3},
\end{equation}
where
\begin{equation}
\alpha= \frac{1}{2} \frac{\text{d} (r^{2} \Omega)}{\text{d} r}.
\end{equation}

\subsection{Numerical implementation}
Rotation is included in the STARS code by adding a new second-order
finite difference equation.  The chemical evolution equations are also
modified to include the new diffusion coefficients and the structure
equations include the $f_{T}$ and $f_{\!p}$ multiplicative
coefficients.

The RoSE code \citep{PotterToutEldridge2012a} was quite unstable,
requiring a number of approximations to ensure stability throughout
the evolution.  The solver often produced negative solutions for
angular momentum in some sections of the star, creating excessive
shear that led to the instability.  To solve this problem, the
perturbed solution used in the Newton-Raphson solver is adapted so that,
at each time step and for every iteration of the solution, the
angular velocity cannot become negative.  This has the effect of
greatly improving the stability and reduces the number of
approximations needed.

Most of the approximation consisted of addition of constant diffusion
coefficients in certain zones, near the core and the surface, to
ensure that sudden changes in behaviour at the boundaries could be
limited.  This was done with functions of the form
\begin{equation}
f(x,x_{0},\sigma)=\frac{1}{2} \left ( 1- \tanh \left (\frac{x-x_{0}}{\sigma} \right ) \right ).
\end{equation}
Because each of these approximations requires three parameters, two for
the function $f$ and a diffusion coefficient, they introduced many
degrees of freedom and required fine tuning to ensure convergence.  We
have eliminated the need for all but one of these.

Other approximations consisted of limiting certain quantities, such as
the amount of shear.  These have all been removed now.  Finally, the
coefficient $D_{\text{con}}$ could be non-zero in non-convective
regions.  This also created instability and was not physically
acceptable.  To avoid this we create a smooth transition to zero for
the diffusion coefficients at the boundary between the radiative zones
and the convective zones.  For example, if we call $D_{\text{con}}'$
the coefficient that is calculated in the whole star, the coefficient
we use is given by
\begin{equation}
D_{\text{con}}=D_{\text{con}}' \times f(\nabla_{\text{r}}-\nabla_{\text{ad}},0,\nabla_{\text{c}}),
\end{equation}
where $\nabla_{\text{c}}$ is a fixed characteristic gradient.  This
cannot be too small because a transition that is too sharp creates
discontinuities that make the code unstable nor too high as to make the
results not physically acceptable.  We use the same method for the
coefficients $D_{\text{shear}}$ and $D_{\text{eff}}$, by multiplying by
$1-f(\nabla_{\text{r}}-\nabla_{\text{ad}},0,\nabla_{\text{c}})$.  With
these modifications, the RoSE code without any magnetic field is much
more stable.

\section{Stellar magnetism}

Previously, most of the models for magnetism in stellar evolution took
magnetic field as a function of the angular momentum in stellar
interior.  A particularity of RoSE and the models presented here is
that the toroidal and poloidal components of the magnetic field are
evolved as independent variables, thanks to two
advection-diffusion equations.  These equations couple rotation and the
two components of the field through the magneto-rotational instability
and a simple $\alpha-\Omega$ dynamo.  Dynamo mechanisms are studied in
the frame of mean field magnetohydrodynamics, developed for example by
\cite{Moffatt1970}, and we follow the method of \cite{Roberts1972} for
the derivation of the dynamo.  In a similar way to the evolution of the
angular momentum, we derive these equations by assuming that
turbulence from rotational and magneto-rotational instabilities leads
to the diffusion of the magnetic field and of the angular momentum, as
well as generation of magnetic field by the dynamo.

\subsection{Magnetic field evolution}

The magnetic field in the interior of a star is coupled to the fluid
movements through magnetohydrodynamics.  First, we once again
assume the velocity field is of the form
\begin{equation}
\label{eq:velocityfield}
\bm{U}=U(r) P_{2}(\cos \theta) \bm{e}_{r}+V(r) \frac{\text{d} P_{2}( \cos \theta)}{\text{d} \theta} \textbf{e}_{\theta} 
\end{equation}
and take $U(r)$ as before (equation \ref{eq:meridionalvelocity}).  In
mean field magnetohydrodynamics, we assume that the fields can be
decomposed into large- and small-scale components.  The
evolution of the large-scale magnetic field is described by the
induction equation
\begin{equation}
\label{eq:Magneticevolution}
\frac{\partial \bm{B}}{\partial t}= \nabla \times \left ( \bm{U} \times \bm{B} \right ) - \nabla \times \left ( \eta \nabla \times \bm{B} \right ),
\end{equation}
where $\eta$ is the magnetic diffusivity.  The magnetic field $\bm{B}$
can be decomposed into a toroidal and a poloidal component,
\begin{equation}
\bm{B}=B_{\phi}(r,\theta) \bm{e}_{\phi} + \nabla \times \left ( A(r, \theta) \bm{e}_{\phi} \right ),
\end{equation}
where $B_{\phi}$ and $A$ give the toroidal and poloidal
fields.  Equations \eqref{eq:velocityfield} and~\eqref{eq:Magneticevolution} become
\begin{equation}
\begin{aligned}
\frac{\partial B_{\phi}}{\partial t} &= r B_{r} \sin \theta \frac{\partial \Omega}{\partial r} + B_{\theta} \sin \theta \frac{\partial \Omega}{\partial \theta} \\
&- \frac{1}{r} \frac{\partial}{\partial \theta} \left ( V(r) \frac{ \text{d} P_{2} ( \cos \theta ) }{\text{d} \theta } B_{\phi} \right ) - \frac{1}{r} \frac{\partial}{\partial r} \left ( r U(r) P_{2} ( \cos \theta ) B_{\phi} \right ) \\
&- \left ( \nabla \times ( \eta \nabla \times \bm{B} ) \right )_{\phi}
\end{aligned}
\end{equation}
and 
\begin{equation}
\begin{aligned}
\frac{\partial A}{\partial t} &= - \frac{2 V(r)}{r} \frac{ \text{d} P_{2} ( \cos \theta ) }{\text{d} \theta} A \cot \theta - \frac{U(r) P_{2} (\cos \theta ) }{r} \frac{ \partial (r A)}{\partial r} \sin \theta \\
&+ \alpha B_{\phi} - \nabla \times (\eta \nabla \times A \bm{e}_{\phi} )_{\phi},
\end{aligned}
\end{equation}
where an $\alpha$ term has been introduced in the last equation to
describe the regeneration of poloidal field by a dynamo.  With the
shellular hypothesis, the term $B_{\theta} \sin \theta \frac{\partial
  \Omega}{\partial \theta}$ vanishes.

We must reduce these equations to one dimension to include them in the
stellar evolution code.  So we choose the $\theta$ dependence of the
magnetic field and perform a suitable latitudinal average of the
evolution equations for its two components.  We
choose $A(r,\theta)=\tilde{A}(r) \sin \theta$ so that, in the purely
diffusive limit, the poloidal field tends towards a dipolar
geometry.  From the imposed form of the magnetic field, we obtain
$B_{r}=\frac{2 \tilde{A} \cos \theta}{r}$ and
$B_{\theta}=-\frac{\text{d} (r \tilde{A} ) }{\text{d} r} \sin
\theta$.  We also want the toroidal field to vanish at the poles to
avoid singularities, so that we choose
$B_{\phi}=\tilde{B}_{\phi}(r)\sin (2 \theta)$.  These choices are not
unique but are the simplest we can make that respect our
requirements.  Finally, we take $\alpha = \tilde{\alpha}(r)$ and
$\eta = \tilde{\eta}(r)$.  We then define a new average of a quantity
$q$ to be
\begin{equation}
< q >_{\text{mag}}= \int_{0}^{\pi/2} q \sin \theta \text{d} \theta.
\end{equation}
Averaging the evolution equations and using the chosen forms for the
components of the magnetic field, we obtain the one-dimensional
equations we need,
\begin{equation} 
\label{eq:magneticfieldevolution1}
\frac{\partial B_{\phi}}{\partial t} = A \frac{\partial \Omega}{\partial r} - \frac{6}{5 r} V B_{\phi} - \frac{1}{10 r} UB_{\phi}+r \frac{\partial}{\partial r} \left  ( \frac{\eta}{r^{4}} \frac{\partial}{\partial r} ( r^{3} B_{\phi} ) \right )
\end{equation}
and
\begin{equation}
\label{eq:magneticfieldevolution2}
\frac{\partial A}{\partial t} = \frac{ 3 V}{2 r} A - \frac{U}{8r} \frac{\partial}{\partial r} (A r)+ \frac{ 8 \alpha}{3 \pi} B_{\phi} + \frac{\partial}{\partial r} \left ( \frac{\eta}{r^{2}} \frac{\partial}{\partial r} ( r^{2} A ) \right ).
\end{equation}
For the boundary conditions, we take the case where diffusion
dominates the other terms, so that $A \propto 1/r^{2}$ and
$B_{\phi} \propto 1/r^{3}$ as $\eta \to \infty$.  This is what we
expect for a dipolar field, so that we assume that $B_{\phi}=0$ and
$B_{\theta} \propto \frac{\partial ( rA)}{\partial r}$ at $r=0$ and
$R$.

\subsection{Angular momentum evolution with magnetic field}

We extend the evolution equation~(\ref{eq:angularmomevolution}) for
the angular momentum using the formulation of \cite{Spruit2002}.  In
this model, the transport of angular momentum is driven by the Maxwell
stress produced by the magnetic field.  This process is assumed to be
diffusive and so the equation becomes
\begin{equation}
\begin{aligned}
\frac{\partial r^{2} \Omega}{\partial t} &=\frac{1}{5 \rho r^{2}} \frac{\partial ( \rho r^{4} \Omega U)}{\partial r} + \frac{3 r}{8 \pi \rho } <(\nabla \times \bm{B} ) \times \bm{B} >_{\phi} \\
&+\frac{1}{\rho r^{2}} \frac{\partial}{\partial r} \left ( \rho D_{\text{tot}} r^{4} \frac{\partial \Omega}{\partial r} \right ),
\end{aligned}
\end{equation}
where the term $D_{\text{tot}}$ now includes magneto-rotational
turbulence as well as rotationally-driven turbulence and convection.
The purely hydrodynamic turbulence comes from shear
instabilities.  We absorb the effective diffusion coefficient
$D_{\text{eff}}$ in $D_{\text{shear}}$.  The diffusion coefficient for
convective transport is $D_{\text{con}}$ and we call $D_{\text{mag}}$
the magnetic diffusion coefficient (section
\ref{subsec:magneticdiffusion}).  The total diffusion coefficient
includes all these effects, $D_{\text{tot}}
=D_{\text{shear}}+D_{\text{con}}+D_{\text{mag}}$.
We have also implicitly chosen $n=2$ so that convective turbulence
leads to a viscous shear that drives convective regions towards
uniform rotation.  There is no good justification for this
\citep[see][]{PotterToutEldridge2012a}.  However differential rotation in the
Sun's convective zone is latitudinal rather than radial and we cannot
model this properly in our one-dimensional shellular model.  Neglect
of convective driving of differential rotation means that we probably
underestimate magnetic field generation in convective zones.
After averaging the
magnetic stress term in the previous equation over $\theta$, we
find
\begin{equation}
\label{eq:AngMomEvolution}
\begin{aligned}
\frac{\partial r^{2} \Omega}{\partial t} &= \frac{1}{5 \rho r^{2}} \frac{\partial ( \rho r^{4} \Omega U)}{\partial r} + \frac{3}{64 \rho r^{3} B_{\phi}} \frac{\partial}{\partial r} \left (r^{3} B_{\phi}^{2} A \right )\\
&+\frac{1}{\rho r^{2}} \frac{\partial}{\partial r} \left ( \rho D_{\text{tot}} r^{4} \frac{\partial \Omega}{\partial r} \right ).
\end{aligned}
\end{equation}

\subsection{Magnetic diffusion}
\label{subsec:magneticdiffusion}

In radiative zones, angular momentum is assumed to be redistributed by
magnetic turbulence driven by the Tayler instability
\citep{Tayler1973}.  We use the turbulent diffusion coefficients of
\cite{MaederMeynet2004}.  The magnetic diffusivity $\eta$ satisfies the
equation
\begin{equation}
\left (N_{T}^{2} + N_{\mu}^{2} \right ) \eta^{2} + \left (2 K N_{\mu}^{2} - \frac{r^{2}\omega_{\rm A}^{4}}{\Omega} \right ) \eta - 2 K r^{2} \omega_{\rm A}^{2} = 0,
\end{equation}
where $N_{T}$ is the Brunt-V\"{a}is\"{a}l\"{a} frequency, $N_{\mu}$ is
the frequency associated with the mean molecular weight gradient, $K$
is the thermal diffusivity and $\omega_{\rm A}$ is the Alfv\`en
frequency.  The magnetic diffusivity is calculated by solving this
equation.

We take the calibration constant $C_{\text{m}}$ of
\cite{PotterChitreTout2012c} to unity.  Finally, we introduce the
chemical Prandtl number $\text{Pr}_{\text{c}}$ and the magnetic
Prandtl number $\text{Pr}_{\text{m}}$ so that the chemical evolution
equation becomes
\begin{equation}
\frac{\partial X_{i}}{\partial t} = \frac{1}{r^{2}} \frac{\partial}{\partial r} \left ( \text{Pr}_{\text{c}} D_{\rm tot} r^{2} \frac{\partial X_{i}}{\partial r} \right ) 
\end{equation}
and $D_{\text{mag}}=\eta/\text{Pr}_{\text{m}}$.  The two Prandtl
numbers describe how efficiently the turbulent diffusivity transports
magnetic flux and chemical composition compared to angular momentum.

\subsection{Dynamo model}

Following \cite{MaederMeynet2004} we describe the $\alpha$-part of the
magnetic dynamo by taking $\alpha=\gamma r /\tau_{\text{a}}$, where
$\gamma$ is an efficiency parameter and $\tau_{a}$ is the
amplification time-scale of the field.  We then use
$\tau_{a}=\frac{N}{\omega_{A} \Omega q}$, where $q=\frac{\partial \log
  \Omega}{\partial \log r}$.  This gives the dynamo efficiency
\begin{equation}
\alpha=\gamma \frac{ r \omega_{A} \Omega q}{N}.
\end{equation}
Following \citet{PotterChitreTout2012c} we choose a fiducial $\gamma =
10^{-16}$ but the question of what choice of $\gamma$ distinguishes between a
super-critical dynamo, that generates magnetic field faster than it
diffusively decays, and a sub-critical dynamo, that does not, is of
interest when the sustainability of poloidal field is important.  We
address this empirically at the start of section~\ref{secpostms} and
deduce that our fiducial choice is significantly super-critical.
We note again that we are modelling neither the latitudinal variation
of the magnetic field nor short time-scale variations.  So we cannot
model magnetic cycles \citep{Roberts1972}.  Nor do we want to model
cycles that have periods much less than the thermal timescale of our star.

\subsection{Magnetic braking}

We include magnetic braking as angular momentum is carried away by
a stellar wind.  Magnetic field spins down the surface of the star
by effectively forcing the wind to corotate to the Alfv\`en radius
\citep{Mestel1987}, at which the magnetic energy density matches the
kinetic energy density in the stellar wind.  We integrate equation
\eqref{eq:angularmomevolution} to get
\begin{equation}
\frac{\text{d} H_{\text{tot}}}{\text{d} t} = 4 \pi R^{4} \rho D_{\text{tot}} \left ( \frac{\partial  \Omega}{\partial r} \right )_{R},
\end{equation}
where $\frac{\text{d} H_{\rm tot}}{\text{d} t}$ is the total rate of
angular momentum loss from the star.  With the formulation of
\cite{WeberDavis1967},
\begin{equation}
\frac{\text{d} H_{\text{tot}}}{\text{d} t} = R_{A}^{2} \Omega \dot{M} = \sigma^{2} J_{\text{surf}},
\end{equation}
where $R_{\text{A}}$ is the Alfv\`en radius,
$\sigma=\text{max}(1,\frac{R_{A}}{R})$ and $J_{\text{surf}}$ is the
specific angular momentum at the surface of the star.  According to
\citet{Doula2002} we can calculate the magnetic efficiency
\begin{equation}
\phi(r)= \frac{B_{\ast}^{2}R^{2}}{\dot{M}V_{\infty}} \frac{ ( \frac{r}{R})^{-4} }{1-\frac{R}{r} },
\end{equation}
where $v_{\infty}=v_{\text{esc}}=\sqrt{2 g_{\text{eff}} R}$ is the
escape velocity at the stellar surface.  The Alfv\`en radius is
typically taken where the dynamo efficiency is equal to one.  Hence,
setting $\phi=1$ and $\sigma=R_{A}/R$, we obtain the equation
\begin{equation}
\label{Equ:MagneticBraking}
\sigma^{4}-\sigma^{3}=\frac{B_{\ast}^{2} R^{2}}{\dot{M} v_{\text{esc}} }.
\end{equation}
Potter et al. only considered the asymptotic cases of either very
strong or very weak magnetic field to determine the interesting
solution for of the equation \eqref{Equ:MagneticBraking} but we solve
this with a Newton method that converges within a few steps.

\subsection{Numerical implementation}

Magnetic effects are implemented in a very similar way to angular
momentum.  Two finite difference diffusion-advection equations are
derived from the evolution equations
\eqref{eq:magneticfieldevolution1}
and~\eqref{eq:magneticfieldevolution2} and the finite difference
equation for angular momentum now includes the new term describing
Maxwell stress.

As for angular momentum, many approximations made by Potter et al. have
been eliminated.  The only diffusion coefficient that needs to be adjusted
is the total diffusion coefficient for the toroidal component of the
magnetic field.  Because of the dipolar structure of the field, the
toroidal component is much stronger near the centre of the star when
the core is fully radiative.  This happens at the end of the main
sequence.  The field can then be higher than $10^{6}\,$G at the
last but one mesh point near the core.  To meet the boundary
condition the field then goes to zero within one mesh point.  This creates a
discontinuity that makes the code unstable.  Hence, we apply a quite
important extra diffusion to the five mesh points closest to the
centre to ensure stability.  Varying this fixed coefficient on the main
sequence has shown that it has no effect on the evolution of the
magnetic field.

Once again, preventing negative angular momentum makes the code much
more stable.  With sufficient precision, it is now possible to simulate
the entire life of most stars given an initial magnetic field and
angular velocity.

\section{Results}

We investigate the origin and evolution of large-scale magnetic fields
in intermediate mass stars.  Because A and~B stars are observed with
various magnetic fields covering a few decades in their surface
intensity, it is interesting to see whether variations in the initial
magnetic fields and rotation can reproduce the diversity of the
observed fields.

\subsection{Free parameters and initial model}

We simulate the evolution of a $3\,\rm M_{\sun}$ star at solar
metallicity $Z=0.02$.  The magnetic Prandtl number is fixed to
$\text{Pr}_{\text{m}}=1$, the chemical Prandtl number is fixed to
$\text{Pr}_{\text{c}}=0.01$ to match the constraints on terminal-age
main-sequence nitrogen enrichment determined by \cite{Hunter2009}.  To
explore the properties of the $\alpha$-dynamo effect we vary the
initial field and the initial stellar spin rate.  The parameter $n$ has a
strong influence on the result, because $n=0$ produces uniform angular
momentum in the convective zone and this results in large shears
throughout these zones.

Given an initial equatorial surface speed $v_{\text{ini}}$, we
generate an initial model that has homogeneous composition and has
reached thermal equilibrium through contraction.  It also has either
uniform angular momentum if $n=0$ or solid body rotation if $n=2$,
such that the surface velocity matches our needs.  As long as they are
reasonable enough to ensure the stability of the code, different
magnetic fields can be added to our model.  Because we expect the field
to be approximately dipolar in the zones where it can be sustained, we
choose the initial fields to behave as $B_{\phi} \propto r^{-3}$ for
the toroidal field and $A \propto r^{-2}$.  The initial strength of the
poloidal field, or more precisely of $A/r$, is chosen to be a few
orders of magnitude smaller than the strength of the toroidal
field.  This choice is motivated by the simulated results, which
indicate that the poloidal component is often much smaller than the
toroidal and this usually leads to numerical stability.

\subsection{Pre-mainsequence evolution}

\begin{figure}
	\begin{center}
		\includegraphics[width=\columnwidth]{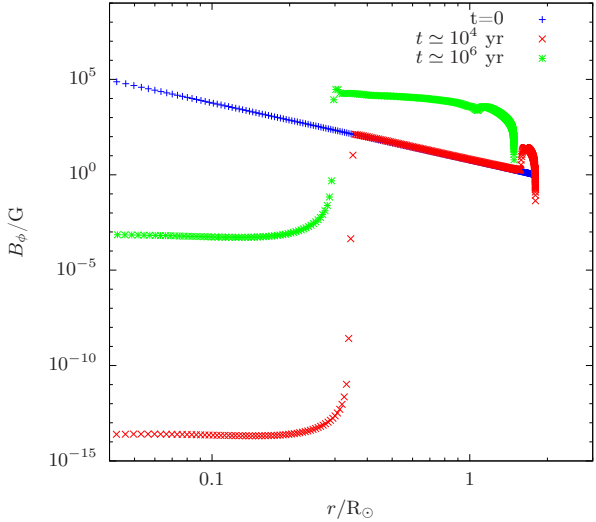}\\
	\end{center}
	\caption{The toroidal field at different times during
          pre-mainsequence evolution.  The large-scale field is completely
          expelled from the convective core and is roughly dipolar in
          the radiative zone.  The drop at the surface is due to the
          boundary conditions.  At $t = 10^6\,$yr the star has reached
          the zero-age main sequence.}
	\label{Figure1}
\end{figure}

\begin{figure}
	\begin{center}
		\includegraphics[width=\columnwidth]{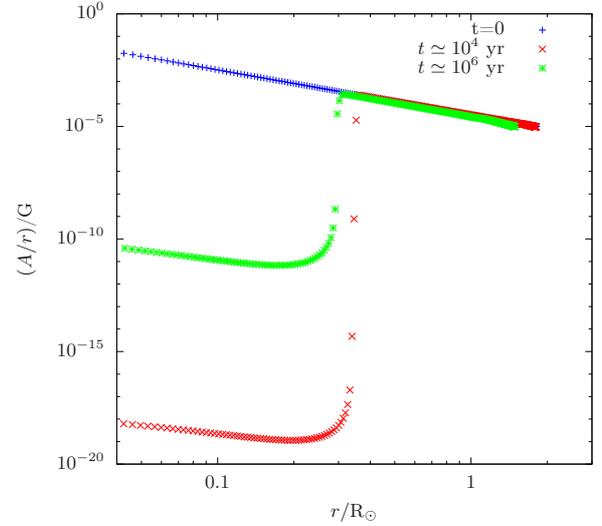}\\
	\end{center}
	\caption{The poloidal field at different times
          during pre-mainsequence evolution.  The large-scale field is also
          completely expelled from the core.  The field behaves as
          $(A/r) \propto r^{-2}$ in the radiative zone, so we expect
          diffusion to be the dominant process there.}
	\label{Figure2}
\end{figure}

Our initial models are $3\,\text{M}_{\sun}$ stars towards the end of
their pre-mainsequence evolution.  They have a radiative envelope and
convective core of approximately $0.7\,\text{M}_{\sun}$ and of radius
$0.35\,\text{R}_{\sun}$, with variations of less than $1\,$per cent
between the non-rotating models and models at half of the critical
rotation rate.  From these models our simulated evolution tracks begin
before any nuclear reactions affect the stellar structure.  This
allows the magnetic field to rearrange itself to its most favoured
form before the beginning of the main sequence.  To illustrate the
field evolution within this period, we simulate the evolution of a
$3\,\rm M_{\odot}$ star, with $n=2$.  Initially, the star is on the
pre-mainsequence, and is in corotation, that is to say with uniform
angular velocity.  The initial surface rotation speed is $100\,\rm
km\,s^{-1}$.  The components of the magnetic field are such that the
field is initially dipolar.  The initial surface strength of the
toroidal component is $1\,\rm G$ and we choose $A/r=10^{-5}\,\rm G$ at
the surface.  The $\alpha$-dynamo efficiency is fixed to
$\gamma=10^{-16}$.  Figs~\ref{Figure1} and~\ref{Figure2} show the two
components at the beginning of the evolution, after approximately
$10{,}000\,$yr and at the beginning of the main sequence,
approximately $10^{6}\,$yr later.

Within the first ten thousand years both components of the large-scale field are
completely dissipated in the convective core.  This is due to the very
large $D_{\text{con}}$ in the convective zones.  In the radiative
zones, the poloidal component still behaves as $B_{\theta} \propto
r^{-2}$.  Because the angular velocity varies throughout the star
creating shear, the toroidal component does not behave exactly as the
poloidal component.  It is roughly dipolar, except near the surface
where magnetic braking results in shears that increase the field,
before it falls to zero within a very narrow region to meet the
boundary condition.  The extra magnetic field is then diffused
throughout the radiative envelope.  The size of this bump increases
with the initial poloidal field because it is sheared into
toroidal.  At the beginning of the main sequence, because the star has
contracted and because of the $\Omega$-dynamo effect, the toroidal
field is much stronger than initially.  However, the poloidal component
is nearly unchanged in its strength, smaller than the toroidal
component by about nine orders of magnitude.  This is because the
$\Omega$-dynamo effect, which converts the shear energy into toroidal
field, operates much more efficiently than the $\alpha$-dynamo effect,
which regenerates the poloidal field.

\subsection{Main-sequence evolution}

A $3\,\rm M_{\odot}$ star takes between 200 and $300\,$Myr to
cross the main sequence and exhaust the hydrogen fuel in its
core.  Low- and intermediate-mass stars spend most of their lives in
this phase.  It is during this period that we expect to see the greatest
attenuation of the field and it is crucial to see whether the initial
field is completely dissipated or not.

We illustrate the main-sequence evolution of the field with similar
models, but with various initial fields.  All the models start in
corotation.  In each case, we use a dipolar initial field and vary its
initial surface strength.  Because of the boundary condition, the
surface toroidal field always vanishes.  As mentioned above, this
boundary condition is met within a very narrow region near the surface
and this is apparent in Fig.~\ref{Figure1}.  Therefore, to describe
the temporal evolution of the toroidal component, we examine a mesh
point situated at $0.99R$.  This is within the very narrow region of
decline but close enough to the surface to allow us to observe
interesting properties.

\begin{figure}
	\centering
	\includegraphics[width=\columnwidth]{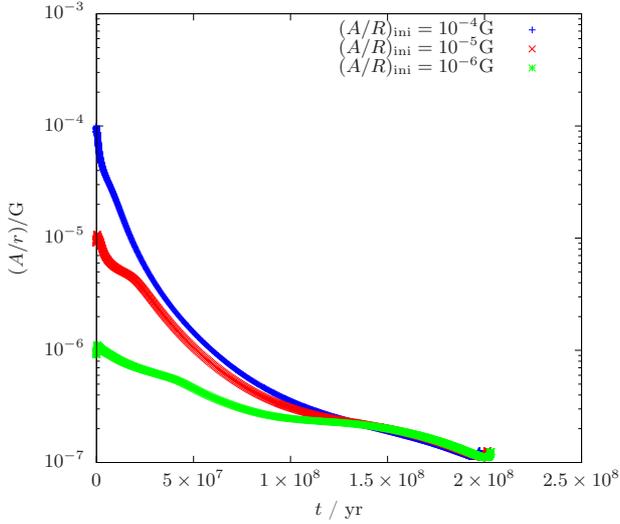}\\
	\caption{Surface poloidal field as a function of time during
          the main sequence.  The initial toroidal field is dipolar and
          its surface strength is fixed at $1.0\,$G. At the end of the
          main sequence, the field seems not to depend much on the
          initial field.}
	\label{Figure3}
\end{figure}

\begin{figure}
	\centering
 	\includegraphics[width=\columnwidth]{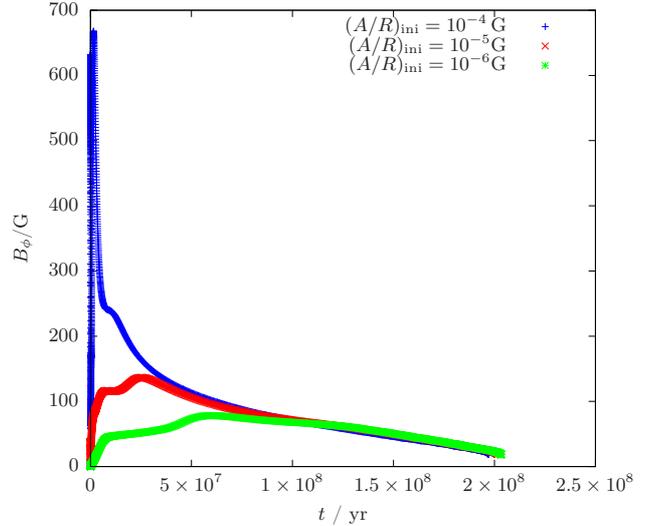}\\
	\caption{Evolution of the toroidal field near the surface as a
          function of time during the main sequence.  Again the initial
          toroidal field is dipolar and its surface strength is fixed
          to $1.0\,$G. Once again, a common field strength is met at the
          end of the main sequence.}
	\label{Figure4}
\end{figure}

\begin{figure}
	\centering
	\includegraphics[width=\columnwidth]{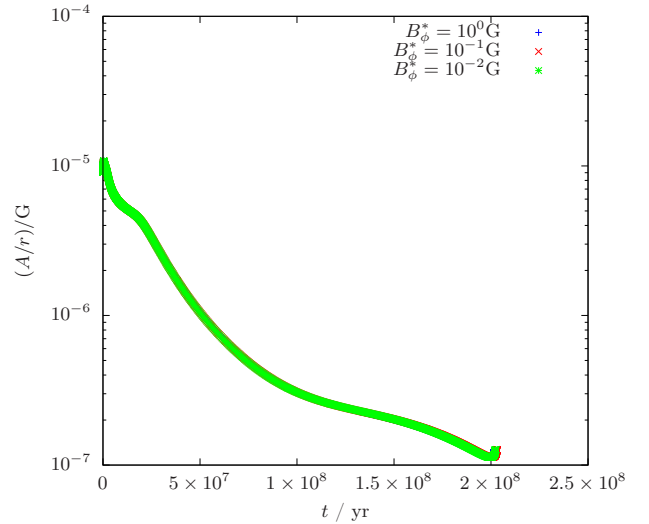}\\
	\caption{Surface poloidal field as a function of time during
          the main sequence.  The initial poloidal field is dipolar and
          its surface strength is reduced to $10^{-5}\,$G.}
	\label{Figure5}
\end{figure}

\begin{figure}
	\centering
	\includegraphics[width=\columnwidth]{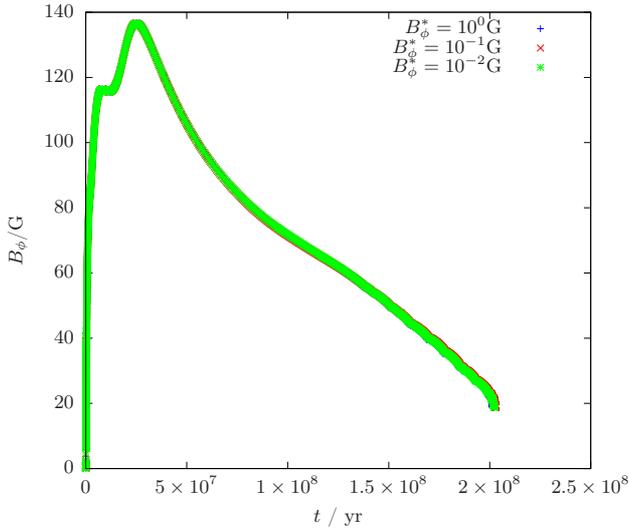}\\
	\caption{Evolution of the toroidal field near the surface as a
          function of time during the main sequence.  The initial
          poloidal field is dipolar and its surface strength is reduced
          to $10^{-5}\,$G.  The initial strength of the toroidal field
          seems to have much less influence on the result than the
          initial strength of the poloidal field.}
	\label{Figure6}
\end{figure}

Figs~\ref{Figure3} to~\ref{Figure6} display the evolution of the two
components of the magnetic field at or near the surface during the
main sequence.  Interesting common features appear when we vary the
initial strength of a component, the other being initially fixed.  One
quite surprising point is that, at the end of the main sequence, each
surface component is very close to what appears to be a common
strength that does not depend on the initial strengths of the
fields.

When the initial poloidal field is strong enough, the toroidal field
reaches a peak at the beginning of the main sequence.  It then
weakens, mainly because the radius of the star increases toward the
end of the main sequence.  For Figs~\ref{Figure3} and~\ref{Figure4} we
vary the initial strength of the poloidal field, the initial toroidal
field being fixed and we do the converse for Figs~\ref{Figure5}
and~\ref{Figure6}.  For both components, the initial poloidal field
seems to have a much larger influence on the early main-sequence
evolution than the initial toroidal field.  Once again, this is linked
to the fact that the $\Omega$-dynamo is more efficient than the
$\alpha$-dynamo at generating toroidal field.

\begin{figure}
	\centering
	\includegraphics[width=\columnwidth]{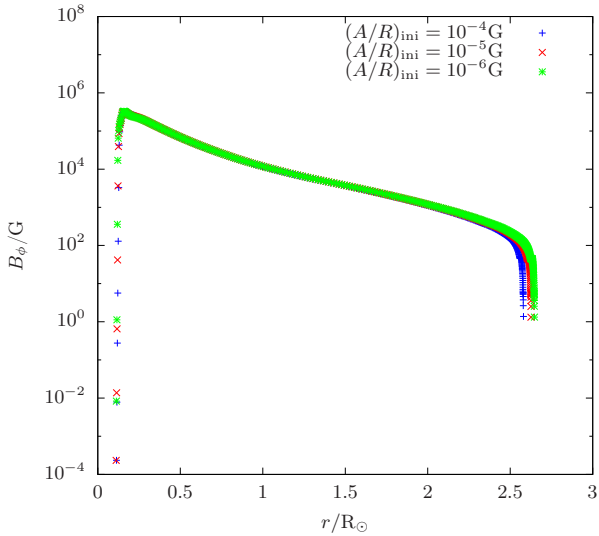}\\
	\caption{Toroidal field in the interior of the star as a
          function of the radius at the end of the main sequence.}
	\label{Figure7}
\end{figure}

\begin{figure}
	\centering
	\includegraphics[width=\columnwidth]{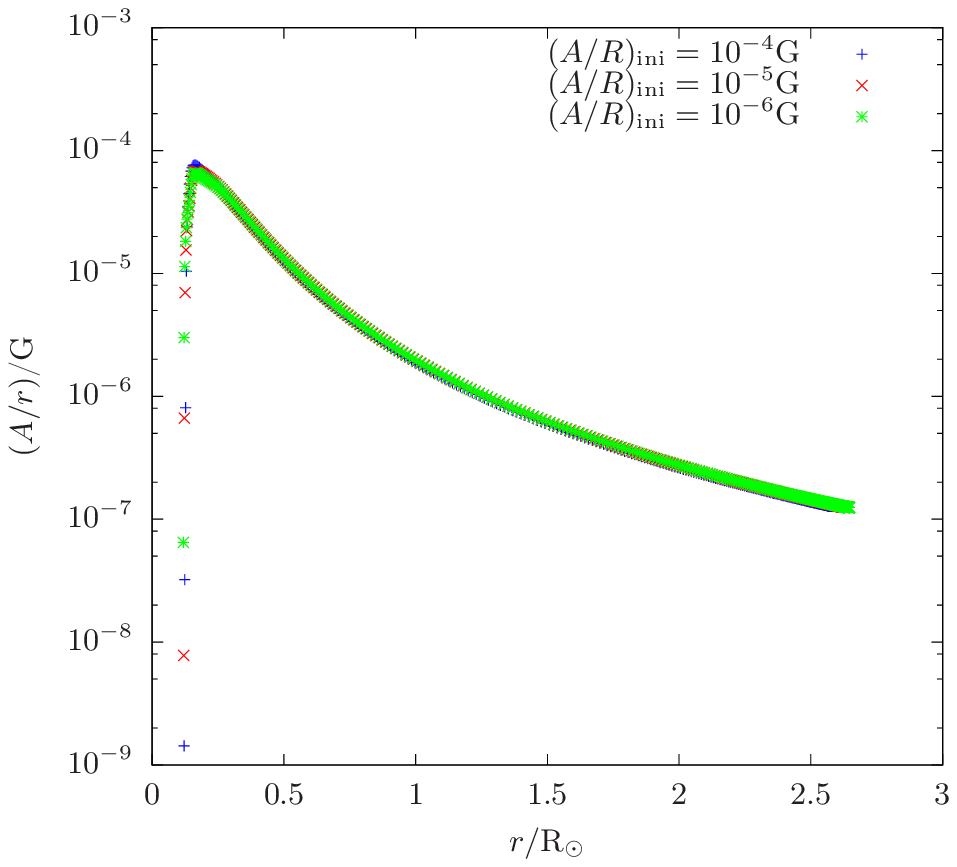}\\
	\caption{Poloidal field in the interior of the star as a
          function of the radius at the end of the main sequence.}
	\label{Figure8}
\end{figure}

\begin{figure}
	\centering
	\includegraphics[width=\columnwidth]{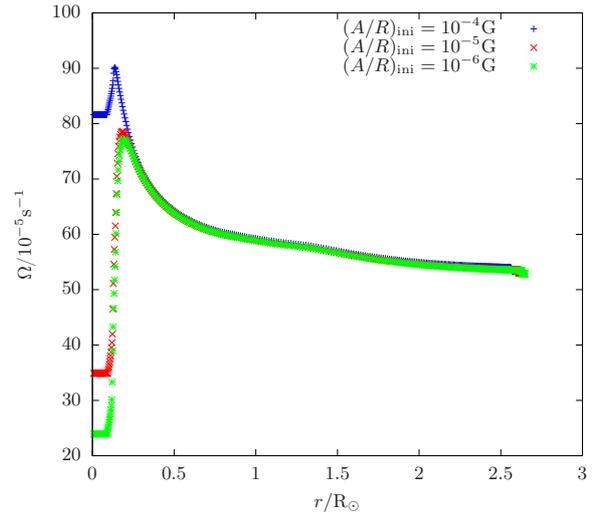}\\
	\caption{Angular velocity in the interior of the star as a
          function of the radius at the end of the main sequence for
          different initial surface poloidal fields.  The drop at the
          surface is due to magnetic braking.}
	\label{Figure9}
\end{figure}

The fact that both surface fields reach strengths that do not depend
on the initial field is quite striking and this raises the same
question for the fields in the interior.  Figs~\ref{Figure7}
and~\ref{Figure8} show the toroidal and poloidal components of the
field at the end of the main sequence, for the same model as before,
with varying strength for initial poloidal field.  We choose a fixed
initial toroidal field because it has much less impact on the
evolution than the initial poloidal field.  Once again, both components
of the field are surprisingly close to each other, despite the initial
poloidal field varying over two decades.  Fig.~\ref{Figure9} shows the distribution
of angular velocity in the interior of the star at the end of the main
sequence for the various initial fields.  The larger the
initial poloidal component, the lower is the angular speed at the end
of the main sequence.  This is due to the fact that the
$\Omega$-dynamo effect is modelled with a term equal to $A
\frac{\partial \Omega}{\partial r}$ in the toroidal field evolution
equation.  Hence, the higher $A$, the more efficiently shear energy
is transferred to the magnetic field.

The fact that the magnetic field does not depend on the initial field
and that the angular velocities in the interior vary significantly
suggests that, given an initial velocity and with a fixed
$\alpha$-dynamo efficiency, there exists an equilibrium for the field
in the radiative zone.  This observation favours the dynamo hypothesis
over the fossil field for stars in the second half of their main
sequence.  Moreover, for stars that are still early on the main
sequence, the initial conditions play an important role.
This conclusion seems to be possibly at odds with the fact that we see
both magnetic and non-magnetic A and~B stars, with surface fields
ranging over a few decades, but so far we have only changed the
initial conditions for the components of the field.

To investigate the effects of initial rotation, we simulate the
evolution of a $3\,\text{M}_{\sun}$ star, with an initial dipolar
field of surface strength $1\, \rm G$ for the toroidal component and
$10^{-5}\, \rm G$ for the poloidal component.  Again, every model
starts in uniform rotation.  The $\alpha$-dynamo efficiency is fixed to
$\gamma=10^{-16}$, we promote solid body rotation with $n=2$ and the
initial surface speed is taken to be $v \in \{ 1,10,20,50,100,150 \}\,
\rm km\,s^{-1}$.  A non-rotating star with the same characteristics has
a radius $R=1.789\, \rm R_{\odot}$ so that, initially, the
various models have $\Omega/\Omega_{\rm crit} \in
\{0.002,0.02,0.04,0.11,0.21,0.33\}$, where $\Omega_{\rm crit}=
\sqrt{\frac{2GM}{3R^{3}}}$ (equation~\ref{Equ:MassLoss}) and
$v$ and $\Omega$ are in the same ascending order.  The
corresponding critical velocity is $v_{\rm crit} = R \Omega_{\rm crit}
\simeq 460\, \rm km\,s^{-1}$.

\begin{figure}
	\centering
	\includegraphics[width=\columnwidth]{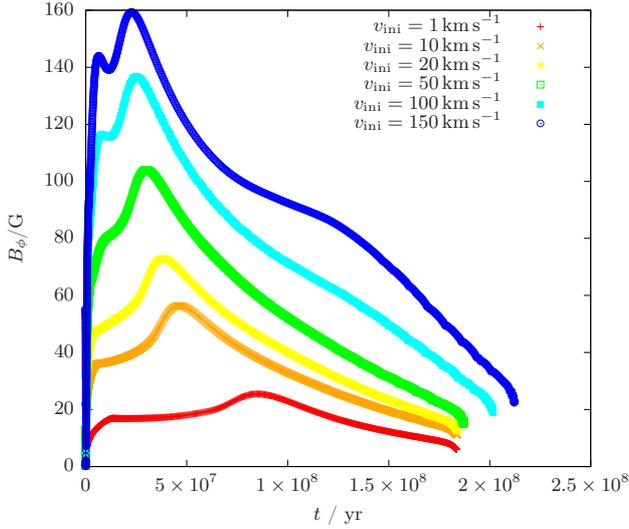}\\
	\caption{Toroidal field strength near the surface as a
          function of time during the main sequence for different
          initial surface speeds.}
	\label{Figure10}
\end{figure}

Except for the lowest velocities, the behaviour of the toroidal field
(Fig.~\ref{Figure10}) is roughly the same as before.  It reaches a peak
at the beginning of the main sequence and then weakens as time
passes.  As expected, the faster the star rotates, the stronger the field
becomes.  The intensity of the field covers approximately two
decades.  In addition, as seen in Fig.~\ref{Figure4},
variations of the initial strength of the poloidal field can produce
variations in the intensity of the surface toroidal field over a few
decades.  Hence, the initial velocity of the star and its initial
poloidal field appear to be the main factors that determine the
intensity of the toroidal field and the combination of the two effects
can produce fields with intensities of a few to over $100\,$G early on.

\begin{figure}
	\centering
	\includegraphics[width=\columnwidth]{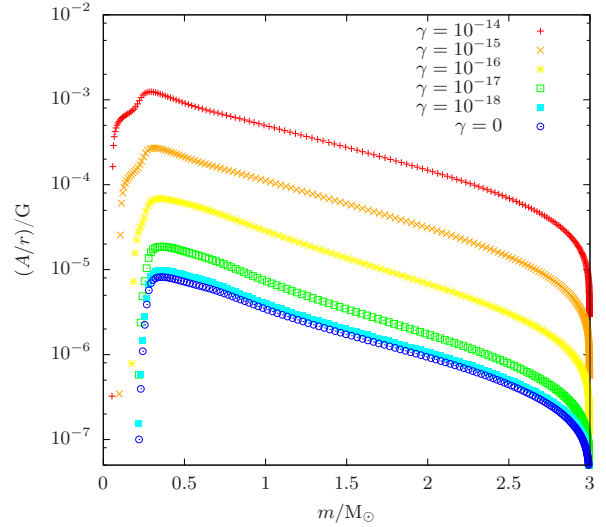}\\
	\caption{Poloidal field in the interior of the star as a
          function of radius at the end of the main sequence for
          various dynamo efficiencies $\gamma$.}
	\label{Figure11}
\end{figure}

\begin{figure}
	\centering
	\includegraphics[width=\columnwidth]{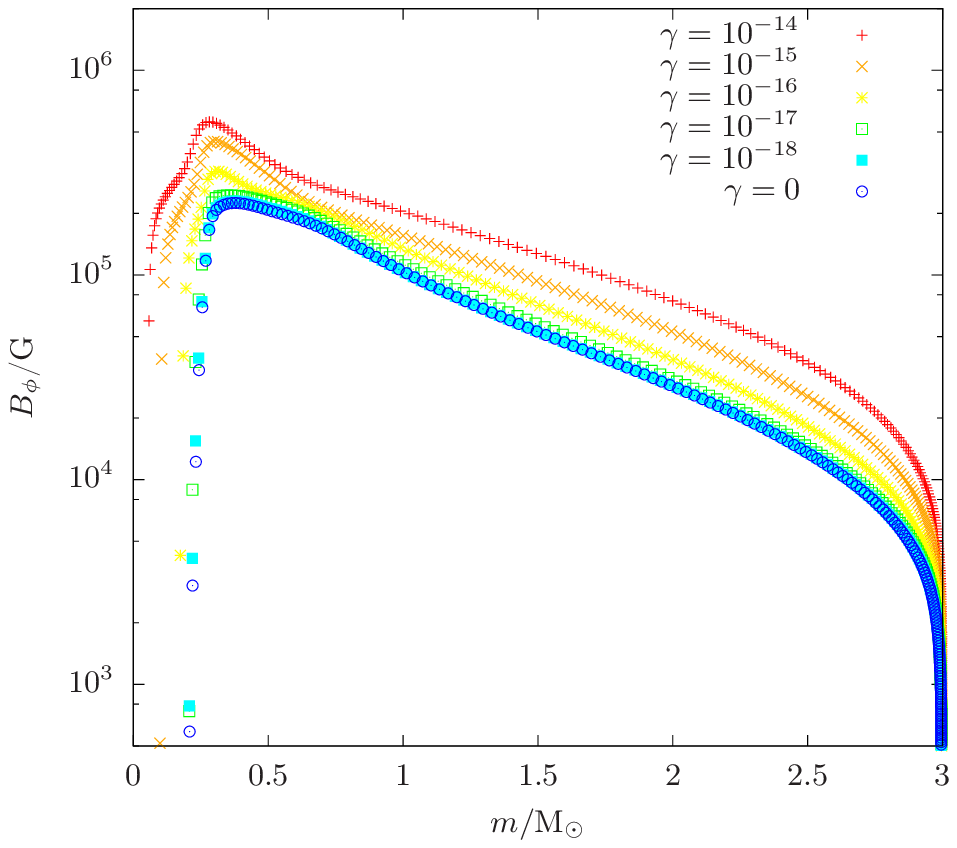}\\
	\caption{Toroidal field in the interior of the star as a
          function of the radius at the end of the main sequence for
          various dynamo efficiencies $\gamma$.}
	\label{Figure12}
\end{figure}

\begin{figure}
	\centering
	\includegraphics[width=\columnwidth]{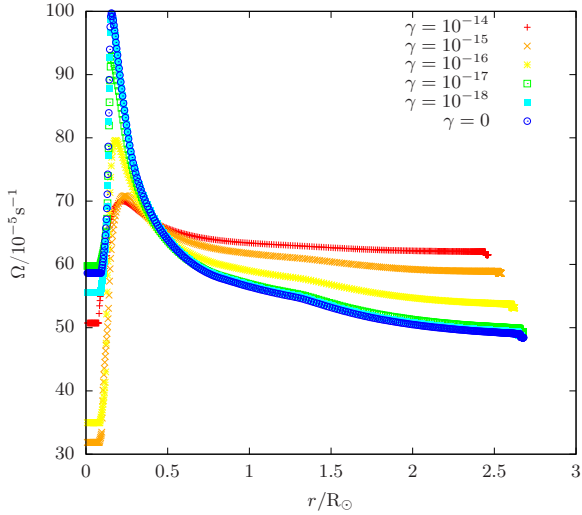}\\
	\caption{Angular velocity in the interior of the star as a
          function of the radius at the end of the main sequence for
          various dynamo efficiencies $\gamma$.}
	\label{Figure13}
\end{figure}
 
\subsection{Post-mainsequence evolution}
\label{secpostms}

The efficiency of the $\alpha$-dynamo does indeed have an influence on
the intensity of the magnetic field over the main sequence.  We
simulate the evolution of the same star as before, with an initial
dipolar field of surface strength $1\,$G for the toroidal component
and $10^{-5}\,$G for the poloidal component and an initial rotation
speed of $100\,\rm km\,s^{-1}$, with the efficiency parameter $\gamma
\in \{10^{-14},10^{-15},10^{-16},10^{-18},10^{-19} \}$.
Figs~\ref{Figure11} to~\ref{Figure13} show the poloidal and toroidal
fields as well as the angular velocity at the end of the main
sequence.  As expected, the poloidal field is stronger for higher
$\alpha$, because more poloidal field is generated for given toroidal.
By the $\Omega$-dynamo, a higher poloidal field also results in a
stronger toroidal field.  Varying $\alpha$ produces greater changes in
the strength of the poloidal field, with variations over more than two
decades for the various $\alpha$s, than in the strength of the
toroidal field, which varies over less than a decade here.  When
$\gamma = 10^{-19}$ the magnetic field evolution is indistinguishable
form when $\gamma = 0$.  So we take this as an indication that
$\gamma\approx 10^{-18}$ is critical and that our fiducial $\gamma =
10^{-16}$ is significantly super-critical.  From
equation~\eqref{eq:AngMomEvolution}, we expect the stronger poloidal
field to enforce more corotation, that is uniform angular velocity.
This is seen in Fig.~\ref{Figure13}, where the models with higher
efficiency are much closer to uniform angular velocity than those with
lower efficiency.

Finally, the magnetic field for $\gamma=10^{-18}$ is nearly the same
as the field with no $\alpha$-dynamo.  This suggests that there exists
a minimum field strength, toward which the field tends when the
$\alpha$-dynamo is very inefficient compared to the other effects
such as $\Omega$-dynamo or diffusion.

After the end of the main sequence and before ending their lives as
white dwarfs, low- and intermediate-mass stars undergo many very
important structural changes.  In the case of a $3\, \rm M_{\odot}$
star, the evolution from the end of the main sequence to the
asymptotic giant branch takes roughly $70\,$Myr, approximately a
third of the time it spends as a main-sequence star.  However, if a
magnetic field can be sustained throughout the main sequence, we might
also expect such a field to be preserved during the late stages of
evolution, or there to be at least some memory of it.

Among the structural changes that are known in stars, some regions see
their convective or radiative behaviour change during the evolution.
This is of particular importance because, as has been discussed for
the pre-mainsequence evolution, convective transport is the main
contributor to the transport of heat, composition, momentum and
magnetic field.  Following the ideas of \citet*{ToutEtAl2004}, we
therefore expect the the magnitude of the averaged large-scale field
to be attenuated in the convective zones compared to the radiative
regions.  They suggested that the dissipation of the field in the
convective zone is due to rapid magnetic reconnection, the conversion
of magnetic energy to kinetic or thermal energy through changes of the
field topology.  We imagine that small-scale rapidly varying magnetic
field is continually generated in convective zones and that the
dissipation of this leads to the enhanced magnetic activity associated
with deep convective envelopes and fully convective stars
\citep{gregory2012}.  This effect prevents the sustenance of any
large-scale field in a convective zone which therefore appears to act
as an insulator.  Though our model does not include any magnetic
reconnection process, the large diffusion coefficient in the
convective zone produces a simple and yet satisfying model for this
dissipation.  Again according to \citet{ToutEtAl2004}, this indicates
that a large-scale field can be sustained as long as the star is partially
radiative.  In the case of a intermediate mass stars, this is the case
during all of the post-mainsequence stages of evolution.

We illustrate this for a $3\,\text{M}_{\sun}$ star, initially in
uniform rotation with an equatorial speed of $100\,\rm km\,s^{-1}$.
The initial field is dipolar, with a surface intensity of $1\,$G for
the toroidal field and $10^{-5}\,$G for the poloidal field. The
$\alpha$-dynamo efficiency is fixed with $\gamma = 10^{-16}$ and
convective zones tend towards solid body rotation.

\begin{figure}
	\centering
	\includegraphics[width=\columnwidth]{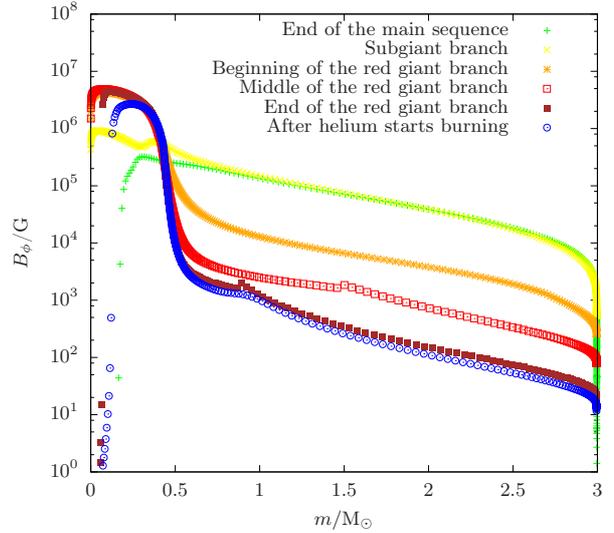}\\
	\caption{Toroidal field inside the star as a function of the
          mass at various times during the sub-giant branch and the
          red giant branch evolution.}
	\label{Figure14}
\end{figure}

After hydrogen fuel has been completely exhausted in the centre, the star
continues to burn in a thin shell around its inert helium core.  During
this phase, the subgiant branch, the core contracts because it is
unable to produce the energy needed to prevent gravitational
collapse.  The core heats as it collapses, increasing the fusion rate
in the hydrogen burning shell.  This leads to expansion of the stellar envelope.  The star
cools slowly and experiences a modest increase in luminosity.
During this phase, the interior becomes fully radiative, so that
magnetic field can penetrate the core.  The yellow points in
Fig.~\ref{Figure14} show the toroidal field during the subgiant branch
phase.

As the star continues to expand, it reaches the red giant branch.  Its
radius can become more than a hundred times what it was in the main
sequence.  It also develops a convective envelope.  The boundary
between the convective zone and the radiative zone sees its depth
increase considerably during this phase.  Because of convection, the
field is slowly attenuated in the outer envelope of the star and tends
to concentrate near the core.  This is illustrated in
Fig.~\ref{Figure14} in which the field is plotted when the star
starts to ascend the red giant branch, approximately $0.3\,$Myr later
and at the end of the red giant phase, just before helium ignites.  At
the end of the red giant phase, the field in the outer parts can be
more than $10^{6}$~times weaker than what is was at the end of the
main sequence.

The end of the red giant phase occurs when the temperature of the core
is high enough to ignite fusion of helium.  At this point, the core of
the star becomes convective.  As on the main sequence, the field is
once again expelled from the core.  Because a convective envelope is
still present, most of the field is then concentrated in a region just
outside the helium burning core, in which the toroidal field can be
lower than $1\,$G, but inside the convective envelope, where the
field can be $10^{5}$ times weaker than in the radiative region.  This
corresponds to the blue points in Fig.~\ref{Figure14}.

After the completion of helium burning in the core, the core starts to
contract again, until it becomes degenerate.  The fusion of helium and
hydrogen continues in thin shells around the core, as for hydrogen in
the subgiant branch.  This phase is the asymptotic giant
branch~(AGB).  Shortly after the core has exhausted its helium fuel, it
becomes radiative again, so that the inner layers of the star,
including the degenerate core, are contained in a radiative core and
the outer regions again become a deep convective envelope.

\begin{figure}
	\centering
	\includegraphics[width=\columnwidth]{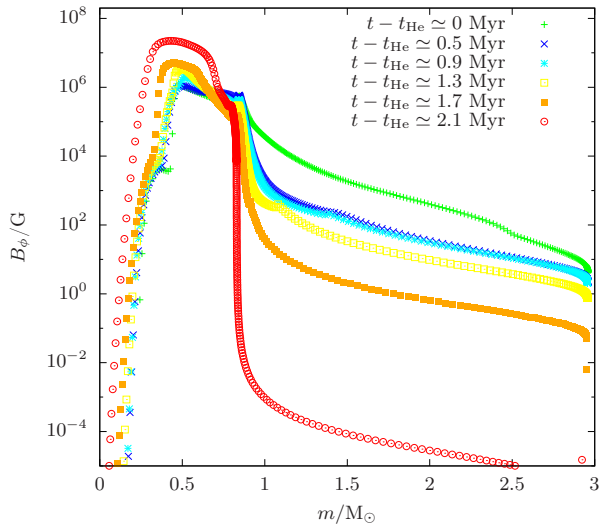}\\
	\caption{Toroidal field inside the star as a function of the
          mass coordinate at various times after helium exhaustion
          in the core, at time $t_{\rm He}$, during the asymptotic
          giant phase.}
	\label{Figure15}
\end{figure}

At the end of the AGB phase, the field is almost completely contained
within the degenerate core of approximately $0.75\,\text{M}_{\sun}$,
the only remaining radiative zone.  The core is extremely compact and
its radius is less than 0.001\,per\,cent of the star but it contains
20\,per\,cent of the total mass of the star.  Hence, we expect the
field to be extremely high in this zone.  In our case, the toroidal
component sees its strength reach between $10^{6}$ and~$10^{8}\,$G.
In the convective zone, the large-scale magnetic fields are extremely
weak: they can be more than $10^{10}$ times weaker than in the core.
This is illustrated in Fig.~\ref{Figure15}, where the toroidal field
is plotted as function of the mass coordinate at various times during
the AGB phase.  As time passes, we see that the field in the envelope
is strongly attenuated, becoming negligible towards the end of this
period.  Once again the field concentrates in the core but without
filling it entirely because the almost superconductivity of very
degenerate material makes it hard for the magnetic field to diffuse to
the centre.  On the other hand, the outer half of the core by mass is permeated with an
extremely strong magnetic field, with strength as high as $10^{8}\,$G.

\subsection{Magnetic field in white dwarfs}

We recall that white dwarfs form at the end of the
lives of low- and intermediate-mass stars.  They are remnants of the
degenerate cores that form in the late stages of stellar
evolution.  The previous example with the $3\,\rm M_{\odot}$ star seems
to support the hypothesis of \citet{ToutEtAl2004} concerning the
formation of very high magnetic field in white dwarfs.  Through the
different stages of its evolution, the studied star never became fully
convective.  This allowed the field to be preserved as it is
transported into the radiative zones.  In particular, during the final
stages of their evolution, intermediate mass stars develop a
convective envelope.  This has the effect of concentrating the magnetic
field in the degenerate core.  Because this degenerate region is
extremely small compared to the original size of the star, magnetic
flux conservation means that the field inside the core becomes
extremely high.  This could be a possible explanation for the very high
magnetic fields observed in some white dwarfs.

Yet, this is inconsistent with the fact that no high-field magnetic
white dwarfs have been observed in wide binaries, as reported by
\cite{ToutEtAl2008}.  If such high magnetic field results from the
isolated evolution of a single star, we would expect to find the same
fraction of high field white dwarfs among binaries as among single
stars.  Because this is not the case, \citeauthor{ToutEtAl2008}
suggested that the high fields are likely to be found only in
cataclysmic variables \citep{Warner1995} or single white dwarfs
derived from merged binary stars.  This is still an open problem that
must be further examined in the future.

White dwarfs have been observed with fields covering many decades,
from non-magnetic to more than $10^{9}\,$G.  Let us now explore a
few possibilities to explain the diversity of the observed
fields.  Before they start their main-sequence evolution, stars less
massive than two solar masses undergo a phase where they become fully
convective.  During that period, we expect the decay of large-scale magnetic field
to be greatly enhanced and we are therefore less likely to observe
magnetic field in these stars.  Because low-mass stars also end up
their lives as white dwarfs, this could explain why non-magnetic white
dwarfs are common.

\begin{figure}
	\centering
	\includegraphics[width=\columnwidth]{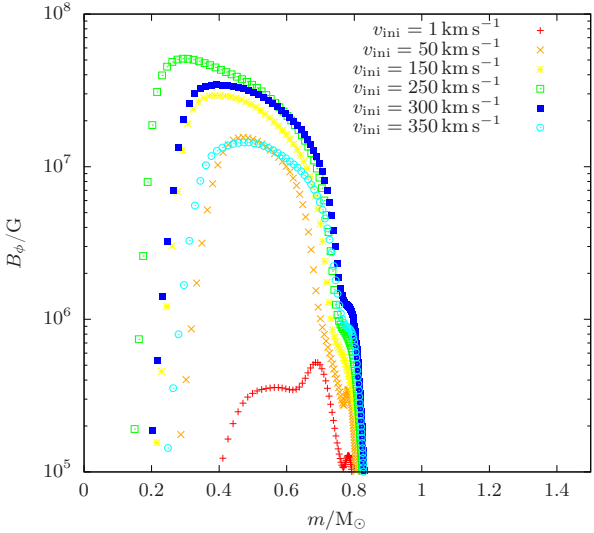}\\
	\caption{Toroidal field in the degenerate core before the
          first thermal pulse as a function of the mass coordinate for
          different initial rotation.  The field in most of the
          envelope has not been represented because it is completely
          negligible.  For initial rotation velocities below
          approximately half of the critical rotation speed, the
          strength of the field increases with the velocity, and for
          velocities above, the strength of the field decreases.}
	\label{Figure16}
\end{figure}

Our code does not allow us to easily produce white dwarfs and the
evolution generally ends in the late stages of the asymptotic giant
branch.  Therefore, to try to evaluate the influence of the magnetic
field during the main sequence on the magnetic field obtained in the
degenerate core at the end of the evolution, we have evolved some of
the different models shown in Fig.~\ref{Figure10}, as well as other
models with the same characteristics but different initial spins, to
the end of their lives.  Their initial equatorial speeds are $v \in \{
1,50,150,250,300,350\}\,\rm km\,s^{-1}$, corresponding to
$\Omega/\Omega_{\rm crit} \in \{0.002,0.11,0.33,0.54,0.65,0.75\}$.  The
toroidal fields obtained in the degenerate core are presented in
Fig.~\ref{Figure16}.

The field does not completely fill the degenerate core but decreases
towards the centre.  As expected, the strength of the field increases
as the initial rotation speed increases, but only for speeds below
approximately half of the critical rotation speed.  For higher angular
velocities, the strength of the field decreases with angular velocity.
Furthermore, the higher the field, the larger the fraction of the core
that is permeated by the field.  We obtained surface toroidal fields
that cover nearly a decade in strength, between $10^{6}$
and~$10^{8}\,$G.  White dwarfs with such fields are considered to be
highly magnetic, even though the initial fields where not very
intense.

\section{Conclusion}
Magnetism and rotation in stars are phenomena that interact strongly.
We have presented a model for the
evolution of angular momentum and magnetic field during the life of a
star that included an $\alpha-\Omega$ dynamo.  The main feature of this
model is to allow the magnetic field to evolve as two independent
components, along with the evolution the angular momentum thanks to
three advection-diffusion equations.

This model, when included in the Cambridge stellar evolution code, has
allowed us derive many results concerning magnetic fields throughout
the life of a $3\, \rm M_{\odot}$ star.  Using various initial
conditions for the magnetic field and the angular velocity, we have
been able to reproduce the diversity of fields observed in
intermediate-mass stars during their main sequence.  We have also
shown that our model seems to favour a dynamo origin rather than a
fossil origin for the magnetic field towards the end of the main
sequence and therefore also for post-main sequence stages of
evolution.

During the late phases of stellar evolution, the rearrangement of the
magnetic field owing to the expansion and contraction of the radiative
and convective zones has confirmed the idea of \citet{ToutEtAl2004}
that, though convective zones in stars appear to act as insulators to
large-scale magnetic fields, they
can be sustained in non-convective regions throughout the life of a
star, as long as it does not become fully convective.  Because this is
the case for intermediate mass stars and because they develop a deep
convective envelope in the very last stage of their evolution on the
asymptotic giant branch, magnetic field concentrates in their
degenerate cores.  The degenerate core is extremely dense and small
compared to a star on its main sequence.  So magnetic flux
conservation leads us to expect a very high field in the core.  Our
numerical simulations have allowed us to reproduce this, giving a
possible explanation for the very high fields observed in magnetic
white dwarfs.

Yet this model is not totally consistent with some observational
evidence.  One open problem is that no highly magnetic white dwarfs
have been observed so far in wide binary star systems, even though the
previous arguments should hold for stars with a companion.  Further
work is required to investigate the effects of binary star interaction
on magnetic field and angular momentum evolution in stars.

\section{Acknowledgements}
LGQ is grateful to the Institute of Astronomy for hospitality during
his stay.  CAT thanks Churchill college for his fellowship.  We thank
the reviewer for helpful comments.




\providecommand{\noopsort}[1]{}



\appendix

%


\bsp	
\label{lastpage}
\end{document}